\documentclass[twocolumn,preprintnumbers,amsmath,amssymb,prl,reprint,groupedaddress,
amsmath,citeautoscript,flushbottom,floatfix,superscriptaddress]{revtex4}

\usepackage{graphicx}% Include figure files
\usepackage{dcolumn}% Align table columns on decimal point
\usepackage{bm}% bold math
\usepackage{amsmath}
\usepackage{amssymb}
\usepackage{amsmath}
\usepackage{booktabs}
\usepackage{multirow}
\usepackage{array}
\usepackage{color}
\usepackage{xcolor}
\usepackage[normalem]{ulem}
\usepackage{float}
\usepackage{tabularx}% Table width
\usepackage{amsmath}% Higher math
\usepackage[colorlinks,linkcolor=blue,citecolor=blue,urlcolor=black]{hyperref}

\makeatletter
\def\@hangfrom@section#1#2#3{\@hangfrom{#1#2#3}}
\makeatother
\newcommand{\vc}[1]{\boldsymbol{#1}}

\begin{document}

\title{Spin-orbit excitons in a correlated metal: Raman scattering study of Sr$_2$RhO$_4$}

\author{Lichen~Wang}
\email[]{lichen.wang@fkf.mpg.de}
\affiliation{Max Planck Institute for Solid State Research, Heisenbergstrasse 1, D-70569 Stuttgart, Germany}
\author{Huimei~Liu}
\email[]{huimei.liu@ifw-dresden.de}
\affiliation{Max Planck Institute for Solid State Research, Heisenbergstrasse 1, D-70569 Stuttgart, Germany}
\affiliation{Institute for Theoretical Solid State Physics and W$\ddot{u}$rzburg-Dresden Cluster of Excellence ct.qmat, IFW Dresden, Helmholtzstrasse 20, 01069 Dresden, Germany}
\author{Valentin~Zimmermann}
\affiliation{Max Planck Institute for Solid State Research, Heisenbergstrasse 1, D-70569 Stuttgart, Germany}
\author{Arvind~Kumar~Yogi}
\affiliation{Max Planck Institute for Solid State Research, Heisenbergstrasse 1, D-70569 Stuttgart, Germany}
\affiliation{UGC-DAE Consortium for Scientific Research (CSR), Indore Centre, University Campus, Khandwa Road, Indore (M.P.) 452001 India}
\author{Masahiko~Isobe}
\affiliation{Max Planck Institute for Solid State Research, Heisenbergstrasse 1, D-70569 Stuttgart, Germany}
\author{Matteo~Minola}
\affiliation{Max Planck Institute for Solid State Research, Heisenbergstrasse 1, D-70569 Stuttgart, Germany}
\author{Matthias~Hepting}
\email[]{hepting@fkf.mpg.de}
\affiliation{Max Planck Institute for Solid State Research, Heisenbergstrasse 1, D-70569 Stuttgart, Germany}
\author{Giniyat~Khaliullin}
\affiliation{Max Planck Institute for Solid State Research, Heisenbergstrasse 1, D-70569 Stuttgart, Germany}
\author{Bernhard~Keimer}
\email[]{b.keimer@fkf.mpg.de}
\affiliation{Max Planck Institute for Solid State Research, Heisenbergstrasse 1, D-70569 Stuttgart, Germany}
\date{\today}

\begin{abstract}
Using Raman spectroscopy to study the correlated 4$d$-electron metal Sr$_2$RhO$_4$, we observe pronounced excitations at 220 meV and 240 meV with $A_\mathrm{1g}$ and $B_\mathrm{1g}$ symmetries, respectively. We identify them as transitions between the spin-orbit multiplets of the Rh ions, in close analogy to the spin-orbit excitons in the Mott insulators Sr$_2$IrO$_4$ and $\alpha$-RuCl$_3$. This observation provides direct evidence for the unquenched spin-orbit coupling in Sr$_2$RhO$_4$. A quantitative analysis of the data reveals that the tetragonal crystal field $\Delta$ in Sr$_2$RhO$_4$ has a sign opposite to that in insulating Sr$_2$IrO$_4$, which enhances the planar $xy$ orbital character of the effective $J=1/2$ wave function. This supports a metallic ground state, and suggests that $c$-axis compression of Sr$_2$RhO$_4$ may transform it into a quasi-two-dimensional antiferromagnetic insulator.

\end{abstract}

\maketitle

Ever since the seminal work of Mott \cite{Mot68}, correlation-driven metal-insulator transitions have been a major focal theme of solid-state research. A multitude of electronic phases -- including unconventional magnetism, charge order, and superconductivity -- have been identified in proximity to the Mott transition of complex materials, and increasingly elaborate models have been devised for their theoretical description \cite{Ima98,Geo96}. Spectroscopy provides some of the most powerful diagnostic abilities of the strength and influence of electronic correlations close to a metal-insulator transition, as epitomized by the single-orbital Hubbard model \cite{Kot04}. In its insulating state, all electrons are bound to atomic sites and can propagate incoherently via excited states with unoccupied and doubly occupied sites. In the metallic state, coherent quasiparticle bands appear at the Fermi level, but the atomic states remain visible as incoherent Hubbard bands \cite{Dam04}. The intensity ratio between coherent and incoherent features in the spectral function can be used to assess the proximity to the metal-insulator transition. Experimental realizations of the single-orbital Hubbard model are rare, however, and in materials for which the Hubbard model is believed to be relevant (such as the cuprate and nickelate superconductors with one hole in the $d$-electron shell), the Hubbard bands tend to overlap with electronic interband transitions, thus confounding spectroscopic studies.

Recent attention has turned to correlated-electron materials with multiple active $d$ orbitals, which are more common and can spawn an even larger variety of electronic phases. The excitation spectra of Mott-insulating systems with multiple active $d$ orbitals are characterized by intra-atomic multiplets generated by the interplay of crystal field, Hund's rule, and spin-orbit interactions; prominent examples are Mott-insulating iridates and ruthenates ~\cite{Yan15,Kim14,Gre20,Fat15}. In metallic systems, the coherent quasiparticles form multiple Fermi surfaces, which are subject to a complex set of instabilities as demonstrated by recent research on iron-based superconductors \cite{Yi17}. Owing to the complex electronic structure of correlated multiband metals, research has largely focused on the ground state and low-energy fermionic excitations near the Fermi level, whereas at least close to a metal-insulator transition, much of the spectral weight is believed to be concentrated in incoherent remnants of atomic multiplet excitations -- analogues of the Hubbard bands in single-orbital compounds.

Here we report the observation of such incoherent spectral features in a carefully selected multiorbital $d$-electron metal. The system we have chosen is Sr$_2$RhO$_4$, a square-planar compound with electronically active Rh ions in the $4d^5$ configuration -- i.e. one hole in the $t_{2g}$ subshell of the $d$-electron manifold in the nearly octahedral crystal field. Prior angle-resolved photoemission (ARPES) and quantum transport experiments have reported multiple Fermi surface sheets with sharp fermionic quasiparticles, as expected for a stoichiometric metal free of any major sources of electronic disorder \cite{Kim06,Bau06,Per06,Hav08,Liu08,Mar11,Ahn15,Zha19}. In contrast to the more widely studied unconventional superconductor Sr$_2$RuO$_4$ (with Ru in $4d^4$ configuration) \cite{Dam00,Mac96,Ber00}, tilt distortions of the RhO$_6$ octahedra narrow the electronic bands, thus effectively enhancing the correlation strength. Conversely, the isoelectronic compound Sr$_2$IrO$_4$ (with Ir in $5d^5$ configuration) is Mott insulating because the larger spin-orbit coupling (SOC) of the $5d$ electrons splits the $t_{2g}$ manifold into a pseudospin $J=1/2$ ground state and $J=3/2$ excited state, thus further narrowing the electronic bands \cite{Kim08}. Owing to the single hole in the $t_{2g}$ manifold, the intra-atomic multiplet excitations of Sr$_2$IrO$_4$ comprise a simple set of $J=1/2 \rightarrow 3/2$ excitations (termed ``spin-orbit excitons'') that are optically inactive but observable by Raman and resonant x-ray scattering spectroscopies~\cite{Yan15,Kim14}. We have used polarization-resolved electronic Raman scattering to detect incoherent but well-defined spin-orbit exciton features in Sr$_2$RhO$_4$, and show that quantitative analysis of the Raman spectra is a rich source of information on the electronic structure of this strongly correlated metal, complementary to prior experiments on the coherent fermionic quasiparticles.

%-------------------------- Fig.1
\begin{figure}
	\centering{\includegraphics[clip,width=8.6cm]{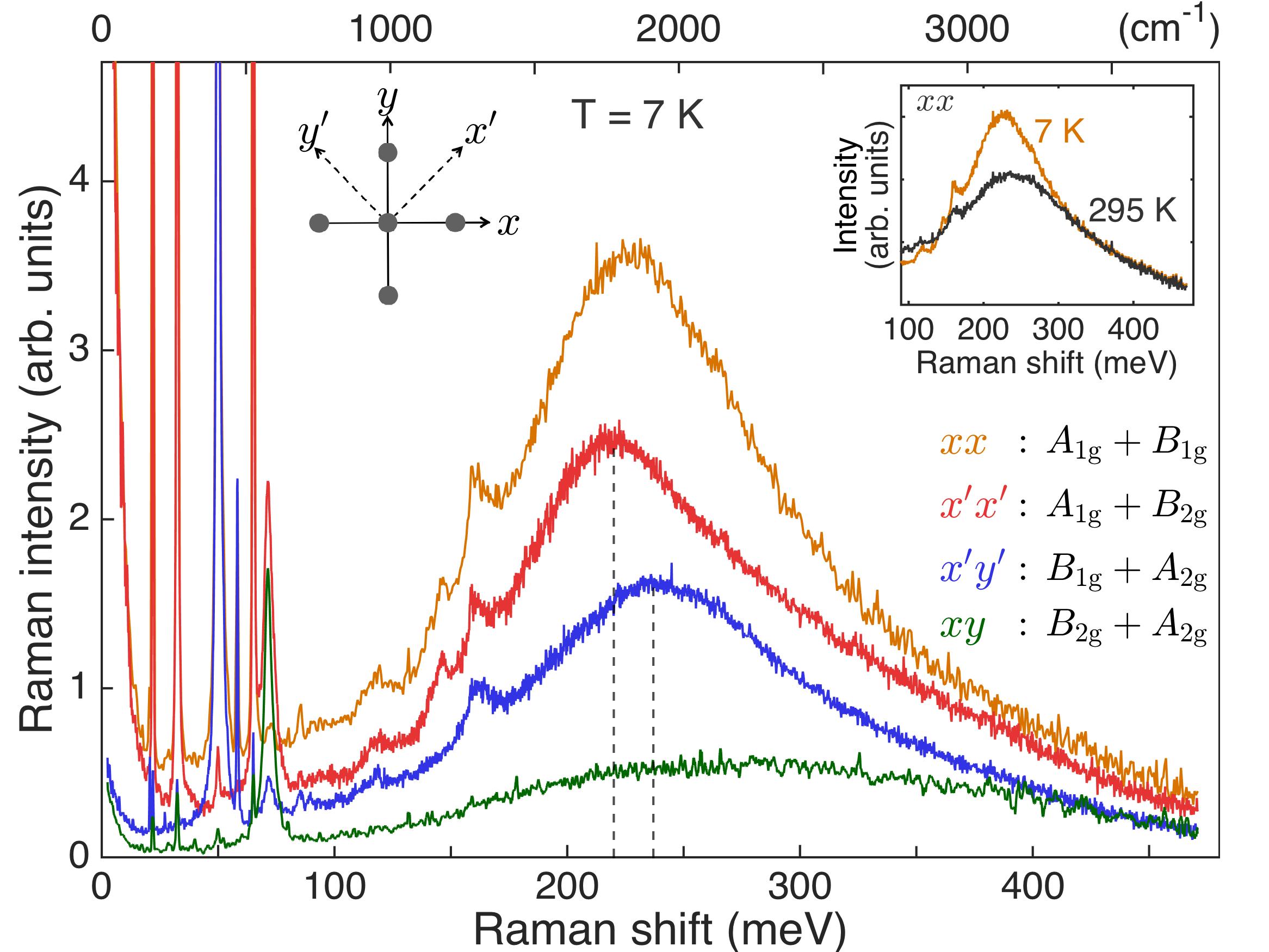}}
	\caption{Raman spectra taken with the 632.8 nm (1.96 eV) excitation line in different scattering geometries. Sharp peaks below 100~meV are phonons. The dashed lines at $220$~meV and $240$~meV indicate the peak positions for broad electronic scattering signals. Small peaks between 100 and 170 meV can be attributed to multiphonons, as in other transition metal oxides (see, {\it e.g.}, Ref. \cite{Ulr06}). Left inset: the light polarization directions with respect to Rh ions (black dots). Right inset: the $xx$ spectra measured at 7 K and 295 K.}
	\label{fig:1}
\end{figure}
%---------------------------

{\it Experiment.}---The single crystals of Sr$_2$RhO$_4$ were grown with the optical floating-zone technique (see the Supplemental Material~\cite{SM} for experimental details). We have collected Raman data covering the energy range up to about 500 meV with four different incident and scattered photon polarizations: $xx$, $x^\prime x^\prime$, $x^\prime y^\prime$, and $xy$, corresponding to the $A_\mathrm{1g}$+$B_\mathrm{1g}$, $A_\mathrm{1g}$+$B_\mathrm{2g}$, $B_\mathrm{1g}$+$A_\mathrm{2g}$, and $B_\mathrm{2g}$+$A_\mathrm{2g}$ representations of the $D_\mathrm{4h}$ point group \cite{Dev07}, respectively. The antisymmetric $A_{2g}$ contribution should be negligible here, as there is no clear sign of time-reversal and/or chiral symmetry breaking in Sr$_2$RhO$_4$. We thus assume that $x^\prime y^\prime$ and $xy$ represent $B_\mathrm{1g}$ and $B_\mathrm{2g}$ spectra, respectively. The observed sharp phonon peaks (Fig.~\ref{fig:1} and \cite{SM}) are consistent with a factor-group analysis and similar to phonon spectra in the isostructural Sr$_2$IrO$_4$~\cite{Gre17}, and are indicative of the high quality of our crystals. Whereas most of the data were taken with the 632.8 nm excitation line of a He-Ne laser, similar spectra were observed with the 532 nm line~\cite{SM}.

\begin{figure}
	\centering{\includegraphics[clip,width=8.6cm]{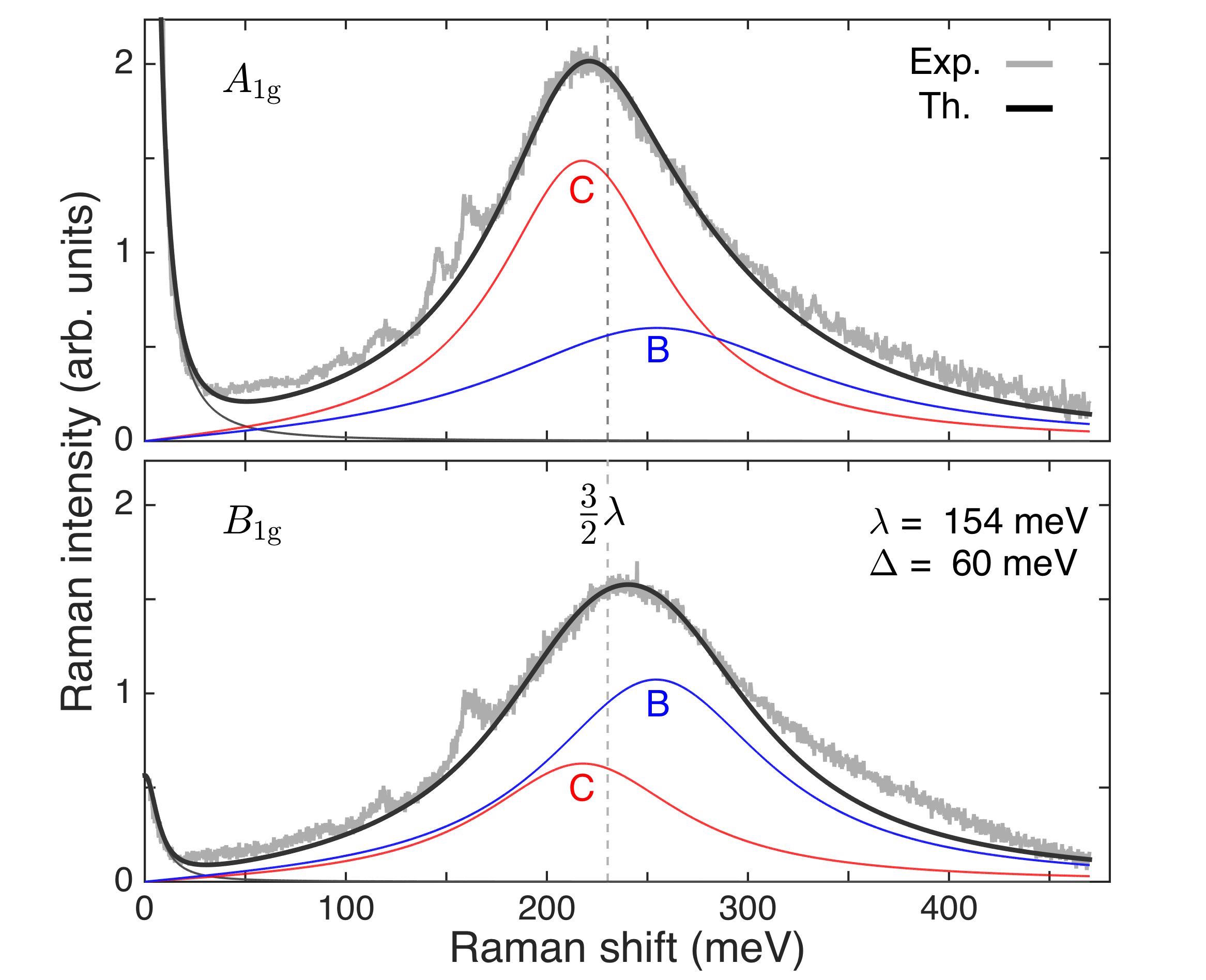}}
	\caption{Experimental spectra in $A_\mathrm{1g}$ and $B_\mathrm{1g}$ channels (gray lines) after extracting the low energy phonon peaks. The fit curves (black) include a Lorenztian tail of the elastic line (thin gray), and the spin-orbit exciton bands $B$ (blue) and $C$ (red). Their peak positions and intensities are calculated theoretically (see text), while the Lorenztian linewidths are adjusted to fit the experimental data.}
	\label{fig:2}
\end{figure}

Figure~\ref{fig:1} shows that the Raman scattering intensity in Sr$_2$RhO$_4$ is dominated by excitations centered around $230$~meV. The peaks are rather broad (the widths are about $150$~meV at 7~K) yet well-defined even at room temperature~\cite{SM}. The $xx$ channel comprising both $A_\mathrm{1g}$ and $B_\mathrm{1g}$ signals exhibits the largest intensity. Since the $xy\!:\!B_\mathrm{2g}$ signal is weak, the $x^\prime x^\prime$ channel $A_\mathrm{1g}$ + $B_\mathrm{2g}$ is dominated by the $A_\mathrm{1g}$ spectra. For the quantitative analysis of the data, we extract the pure $A_\mathrm{1g}$ spectra from the raw data in two different ways, as $xx-x^\prime y^\prime$ and $x^\prime x^\prime-xy$, which yield nearly identical results~\cite{SM}. Their average is presented in Fig.~\ref{fig:2}, along with the $x^\prime y^\prime\!:\!B_\mathrm{1g}$ spectra. Compared to the $B_\mathrm{1g}$ symmetry, the $A_\mathrm{1g}$ signal is stronger and its peak position is lower by about 20 meV.

%------------------Fig. 2
%-------------------

The spectra in Figs.~\ref{fig:1} and \ref{fig:2} are highly unusual for clean metals with well-defined quasiparticles~\cite{Per06}, which typically only exhibit a featureless electronic continuum (as exemplified by Sr$_2$RuO$_4$ \cite{Yam96,Phi21}.) A magnetic (e.g., two-magnon~\cite{Dev07,Gre16}) origin of the strong Raman scattering at $230$~meV is unlikely, as the largest intensity is seen in the $A_\mathrm{1g}$ channel with parallel polarization. On the other hand, Raman features with closely similar line shapes were observed in the Mott insulators Sr$_2$IrO$_4$~\cite{Yan15} and $\alpha$-RuCl$_3$~\cite{War20,Lee21}, and assigned to transitions between the intra-ionic $J = 1/2$ and $J = 3/2$ states \cite{Kim14}. These spin-orbit excitons have been found to persist also in lightly doped metallic iridates~\cite{Cla19, Gre16} (although in this case, doping-induced disorder complicates the interpretation of the spectra). Moreover, the energy scale of $230$~meV in Sr$_2$RhO$_4$ is very similar to the spin-orbit splitting $\frac{3}{2} \lambda$ with $\lambda \sim 160$~meV (slightly reduced from a Rh$^{4+}$ free-ion value of $\sim 190$~meV~\cite{Abr70} by covalency effects). Remarkably, spin-orbit levels in the range of $200-250$~meV  have been predicted by quantum chemistry calculations for Sr$_2$RhO$_4$~\cite{Kat14}. The above considerations have led us to develop a theory for Raman scattering from spin-orbit excitons based on a localized model, which describes most aspects of our data on a quantitative level. Note that the theory does not address the width of the excitonic profiles, which is slightly larger in Sr$_2$RhO$_4$ than in Sr$_2$IrO$_4$~\cite{Yan15} and $\alpha$-RuCl$_3$~\cite{War20}, likely due to interaction with fermionic quasiparticles.

\begin{figure}
	\centering{\includegraphics[clip,width=8.6cm]{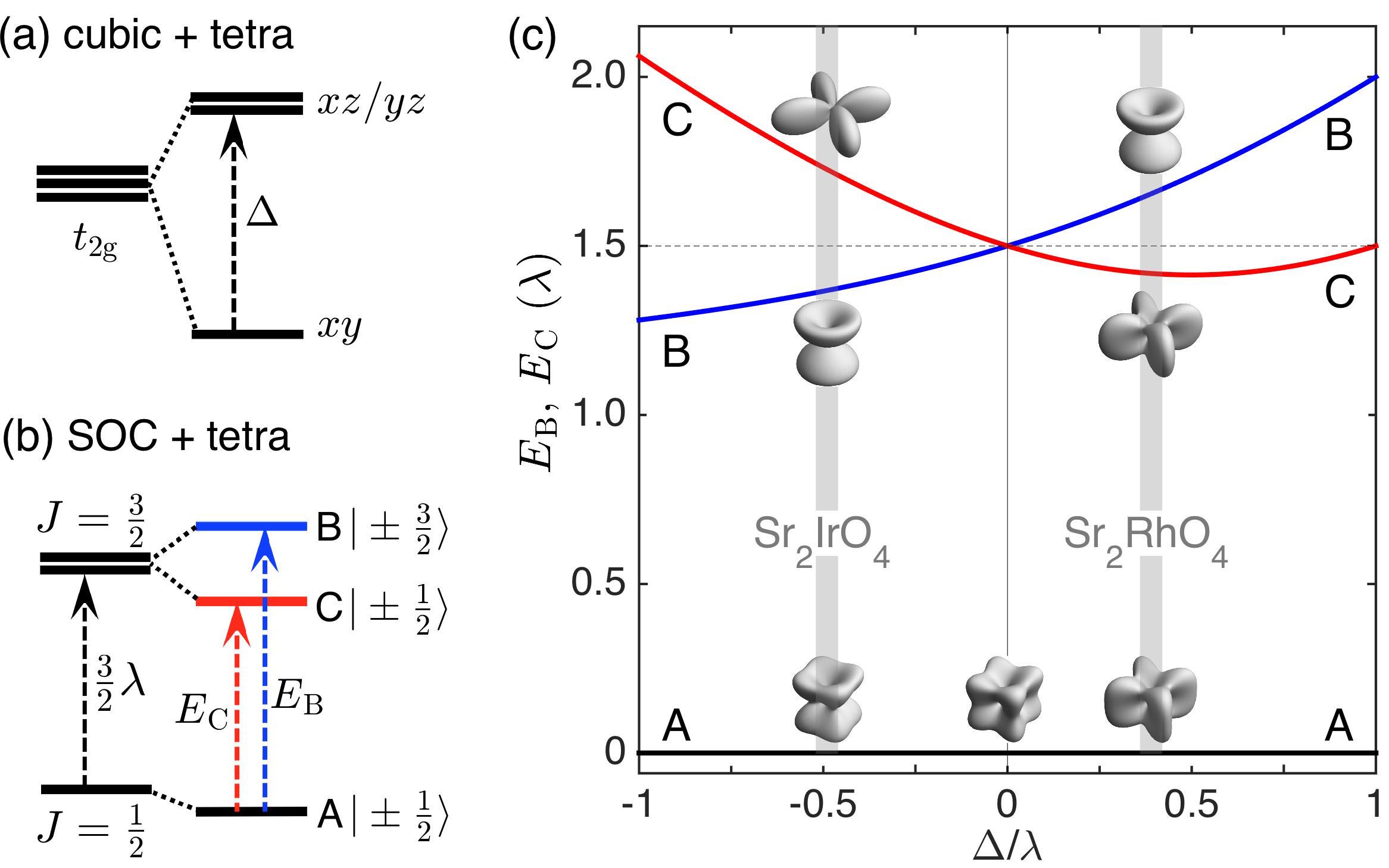}}
	\caption{Energy levels for the $t_\mathrm{2g}$ hole in tetragonal crystal field (a) without and (b) with spin-orbit coupling. In the hole language used here, $\Delta>0$ corresponds to the oxygen octahedra with longer $z\!\parallel \!c$ axis as in Sr$_2$RhO$_4$. (c) Excitation energies $E_\mathrm{B}$ and $E_\mathrm{C}$ as a function of $\Delta/\lambda$. The vertical gray stripes at $\Delta/\lambda \sim 0.4$ and $\Delta/\lambda \sim -0.5$ correspond to Sr$_2$RhO$_4$ with $\lambda \simeq 154$ meV and $\Delta \simeq 60$ meV (this work), and Sr$_2$IrO$_4$ with $\lambda \simeq 380$ meV and $\Delta \simeq -190$ meV (Ref.~\cite{Kim14}), respectively. The spatial shapes of the $t_\mathrm{2g}$ hole wave functions in these two compounds are displayed; note a more flat, $xy$-type shape of the ground-state wave function in Sr$_2$RhO$_4$, in contrast to more out-of-plane $xz/yz$ character in Sr$_2$IrO$_4$.}
       \label{fig:3}
\end{figure}

{\it Theory.}---In an octahedral crystal field, the Rh$^{4+}$ ion contains a single hole in the $t_\mathrm{2g}$ orbital level, hosting spin $s=1/2$ and effective orbital $l=1$ moments. The crystal field induced by the tetragonal distortion, $\frac{1}{3}\Delta(n_{yz}+n_{zx}-2n_{xy})= \Delta(l_z^2-\frac{2}{3})$, splits this level into an $xy$ singlet and an $xz/yz$ doublet (Fig.~\ref{fig:3}(a)). On the other hand, the spin-orbit coupling $\lambda (\bm{l \cdot s})$ forms multiplets with total angular momentum $J=1/2$ and $3/2$. The combined action of these interactions results in a level structure comprising the ground state Kramers doublet A with an effective spin $1/2$, and two excited doublets B and C derived from the $J=3/2$ quartet states with $J_z=\pm 3/2$ and $J_z=\pm 1/2$, respectively, see Fig.~\ref{fig:3}(b). Explicit forms of the corresponding spin-orbit entangled wave functions can be found in the Supplemental Material~\cite{SM}, and their spatial shapes at representative $\Delta/\lambda$ values are illustrated in Fig.~\ref{fig:3}(c). With respect to the ground state A level, the energies of B and C doublets read as
\begin{align}
  E_\mathrm{B}=\frac{3}{4}\lambda+\frac{1}{2}\Delta+\frac{1}{2}R\;, \ \ \ \ \ \ \ \ \ \   E_\mathrm{C}=R \;,
 \label{eq:E}
\end{align}
where $R=\sqrt{\frac{9}{4}\lambda^2+\Delta^2-\lambda\Delta}$. For $\Delta>0$ ($\Delta<0$), the B level is higher (lower) than the C level, see Fig.~\ref{fig:3}(c).

%-------------------Fig.3
%----------------

As mentioned above, the spin-orbit excitons interact with the underlying electronic continuum, so that their spectral features are broadened. In the following, we focus on the polarization dependence of the exciton peak energies and intensities, leaving the line shape effects aside. To this end, we adopt the Fleury-Loudon theory, which describes the Raman scattering intensity in terms of spin exchange operators, with proper form factors encoding the scattering geometry~\cite{Fle68,Dev07}.

\begin{figure}
	\centering{\includegraphics[clip,width=8.6cm]{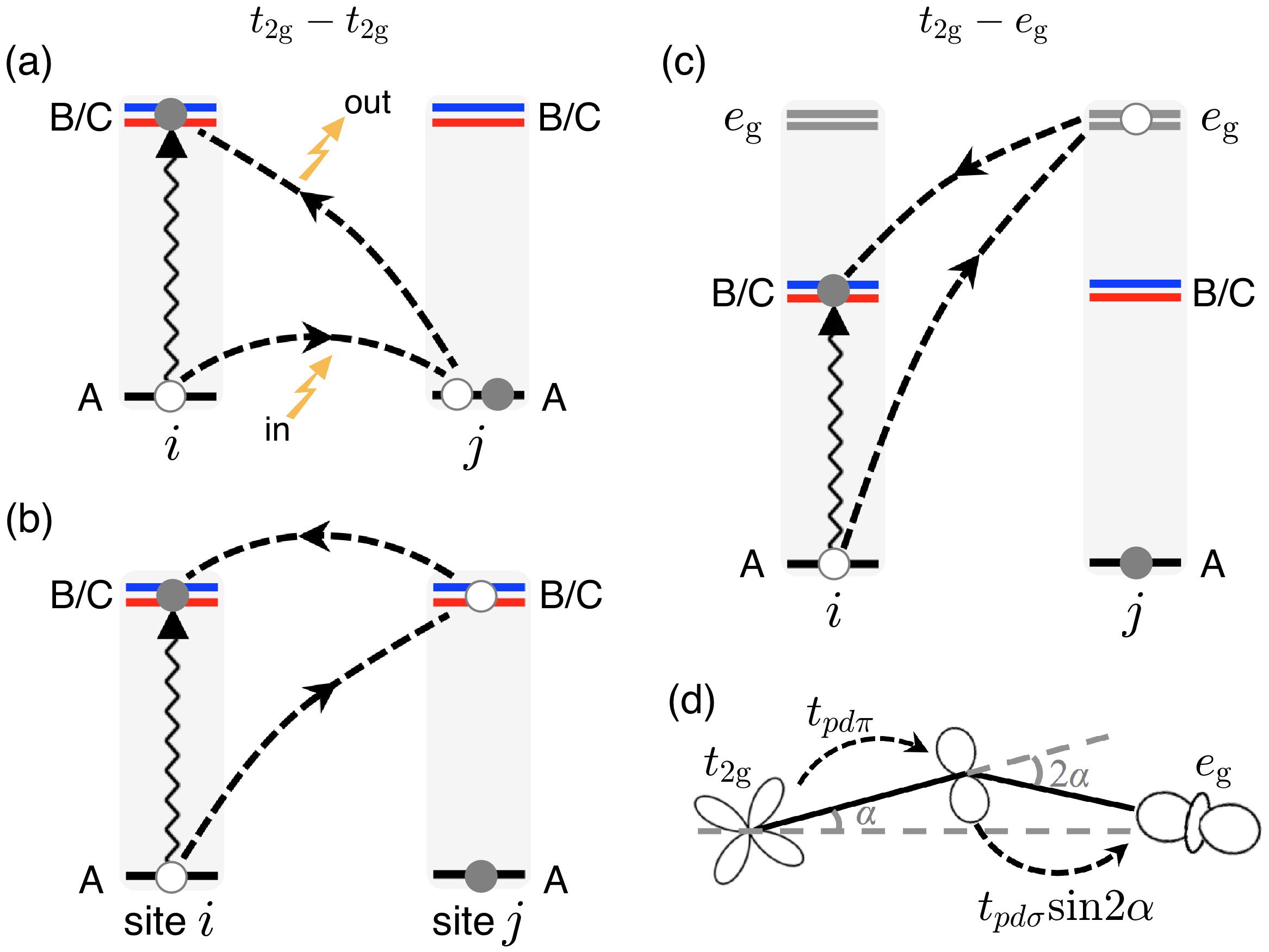}}
	\caption{Schematic of (a)-(b) $t_\mathrm{2g}$-$t_\mathrm{2g}$ and (c) $t_\mathrm{2g}$-$e_{\rm g}$ intersite hoppings (dashed lines) in the Raman photon-in photon-out (orange zigzag arrows) scattering process, creating spin-orbit excitations A$\rightarrow$ B/C (wavy lines) on site $i$. In the intermediate virtual state, site $j$ is occupied by two holes (filled and empty circles) which experience intraionic Coulomb interaction (not shown). (d) $t_{\rm 2g}$-$e_{\rm g}$ hopping between $d_{xy}$ and $d_{3x^2-r^2}$ orbitals, generated by the rotations of the oxygen octahedra around $z\parallel c$ axis by angle $\alpha$. In Sr$_2$RhO$_4$, $\alpha = 10.5 ^{\circ}$~\cite{Sub94}.
}
       \label{fig:4}
\end{figure}

In the present context, the Raman light scattering involves two subsequent optical transitions of $d$ electrons between neighboring Rh ions, and creates a spin-orbit exciton in the final state, as illustrated in Fig.~\ref{fig:4}. As exactly the same intersite hoppings via doubly occupied intermediate states also appear in the derivation of the spin-orbital exchange interactions between ions, the Raman scattering operator $\mathcal{R}$ can be expressed via the corresponding exchange Hamiltonian. Based on this observation~\cite{Fle68}, we can write down the Raman operators of $A_\mathrm{1g}$ and $B_\mathrm{1g}$ symmetries as $\mathcal{R}_{A_{\rm 1g}/B_{\rm 1g}}=(\mathcal{R}_x\pm\mathcal{R}_y)$, where $\mathcal{R}_\gamma$ with $\gamma\in\{x,y\}$ is the spin-orbital exchange operator acting on nearest-neighbor $\langle ij \rangle\parallel \gamma$ bonds. For the $t_\mathrm{2g}$ orbital systems with spin one-half, it has the following structure (neglecting Hund's coupling corrections)~\cite{SM,Kha03}:
\begin{equation}
  \mathcal{R}_\gamma \propto\! \!\sum_{\langle ij\rangle_\gamma} \!\!\left[
  (4 \bm{s}_i\!\cdot\!\bm{s}_j\!+\!1) \mathcal{O}^{(\gamma)}_{ij}\!-\! l^2_{\gamma i}-\! l^2_{\gamma j} +\tau^2 (l^2_{zi}\!+\!l^2_{zj}) \right]\!,
\label{eq:R}
\end{equation}
where the orbital operator $\mathcal{O}^{(x)}_{ij}= [(1-l^2_y)_i(1-l^2_y)_j+(l_y l_z)_i(l_z l_y)_j]+[y\leftrightarrow z]$, and $\mathcal{O}^{(y)}_{ij}$ follows by symmetry (replacing $y \rightarrow x$). The orbital angular momentum operators $l_x=i(d_{xy}^{\dag} d_{zx}-d_{zx}^{\dag} d_{xy} )$, etc.

In Eq.~\eqref{eq:R}, all terms but the last originate from the $t_{\rm 2g}$-$t_{\rm 2g}$ orbital hoppings~\cite{Kha03}, while the $\tau$ term stands for a contribution from nondiagonal $t_{\rm 2g}$-$e_{\rm g}$ hopping $\tilde{t}$. Relative to the $t_{\rm 2g}$-$t_{\rm 2g}$ hopping $t$, it is given by $\tau\equiv \tilde t/t=\sin 2\alpha \;(t_{pd\sigma}/t_{pd \pi})$, where $\alpha$ quantifies the deviation of the Rh-O-Rh bond angle from the ideal $180^\circ$, see Fig.~\ref{fig:4}(d). This contribution can be sizeable even for small $\alpha$, due to the stronger $\sigma$-type overlap $t_{pd\sigma}$ between O-$p$ and Rh-$e_g$ orbitals, as compared to the $t_{pd \pi}$ overlap for $t_{2g}$ states.

%-------Fig.4---------
% ---------------------

In general, the operator $\mathcal{R}_\gamma$~\eqref{eq:R} creates spin, orbital, and composite spin-orbital excitations in the Raman spectra. To proceed further, we need to express this operator in terms of the A, B, and C Kramers doublet states. The mapping onto the spin-orbit basis is straightforward (the details are provided in the Supplemental Material~\cite{SM}). As a result, we obtain the Raman operators of the $A_\mathrm{1g}$ and $B_\mathrm{1g}$ symmetries, and evaluate their matrix elements corresponding to the transitions from the ground state A to the excited states B and C. Neglecting nonlocal spin correlations in paramagnetic Sr$_2$RhO$_4$, we arrive at the following intensities of the B and C excitons in the $A_\mathrm{1g}$ scattering channel:
\begin{flalign}
A_\mathrm{1g}\; & \left \{ \begin{array}{l}
I_\mathrm{B}=\tfrac{1}{4}c^6_{\theta} + \tfrac{1}{2} c^2_{\theta} (1+s^2_{\theta})^2,
\\ [2mm]
I_\mathrm{C}= \tfrac{3}{4} s^2_{\theta}c^2_{\theta} (1+s^2_{\theta})^2 + s^2_{\theta} c^2_{\theta}(c^2_{\theta}+4 \tau^2)^2.
\label{eq:A1g}
\end{array}
\right .
\end{flalign}
In the $B_\mathrm{1g}$ channel, we obtain:
\begin{flalign}
&\ \;\;\;\; B_\mathrm{1g}\;  \left \{ \begin{array}{l}
I_\mathrm{B}=\tfrac{5}{4}c^6_{\theta} + \tfrac{1}{2} c^2_{\theta}(1+s^2_{\theta})^2 ,
\\ [2mm]
I_\mathrm{C}= \tfrac{3}{4} s^2_{\theta}c^2_{\theta} (1+s^2_{\theta})^2.
\label{eq:B1g}
\end{array}
\right . &
\end{flalign}
Here, $c_{\theta} \equiv \cos\theta$, $s_{\theta} \equiv \sin\theta$, and the angle $0<\theta<\frac{\pi}{2}$ given by $\tan 2\theta = 2\sqrt{2}\lambda/(\lambda-2\Delta)$ depends on the relative strength of the tetragonal field $\Delta/\lambda$. It decides the spatial shapes of the spin-orbit wave functions shown in Fig.~\ref{fig:3}(c), and thus determines the intersite hopping amplitudes and Raman matrix elements. The $\tau$ term contributes only to the $A_\mathrm{1g}$ intensity.

We note that within the present nearest-neighbor hopping model, the spin-orbit excitons do not contribute to the $B_\mathrm{2g}$ scattering. In reality, however, a small contribution $\propto (t'/t)^4$ to the $B_\mathrm{2g}$ signal is expected (and possibly present in Fig.~\ref{fig:1}) due to longer-range hoppings $t'$.

Equations ~\eqref{eq:E}, \eqref{eq:A1g}, and~\eqref{eq:B1g} fully determine the spin-orbit exciton peak positions and intensities in the Raman spectra, as a function of three parameters: $\lambda$, $\Delta$, and $\tau$. We recall that $\tau$ accounts for the $t_\mathrm{2g}$-$e_g$ hopping, and its value is determined by the $t_{pd\sigma}/t_{pd \pi}$ ratio (for a given angle $\alpha$).

{\it Discussion.}---The above theory with $\lambda=154$~meV, $\Delta=60$~meV, and $t_{pd\sigma}/t_{pd \pi}=1.2$ reproduces the experimental spectra (Fig.~\ref{fig:2}) very well. The deviations can be attributed to the particle-hole continuum, possible exciton-phonon sidebands, and multiple excitons. Regarding the latter, we note that multiple exciton creation requires at least two subsequent intersite hoppings between the $J=1/2$ and $J=3/2$ levels, which are small in compounds with $180^\circ$ exchange bonding. Therefore, the double-exciton peaks are not clearly visible in the perovskites Sr$_2$RhO$_4$ and Sr$_2$IrO$_4$. In contrast, such intersite hoppings are strong in Kitaev materials with $90^{\circ}$ bonding, and pronounced double-exciton peaks have indeed been observed in $\alpha$-RuCl$_3$~\cite{War20,Lee21}.

Both B and C excitations feed into the $A_\mathrm{1g}$ and $B_\mathrm{1g}$ channels, but with unequal spectral weights (Fig.~\ref{fig:2}). The overall peak position thus depends on the scattering geometry. The $A_\mathrm{1g}$ peak is more intense than $B_\mathrm{1g}$ due to $t_\mathrm{2g}$-$e_g$ hopping; the opposite is expected in Ba$_2$IrO$_4$ with no octahedral rotations.

The relative order of the $A_\mathrm{1g}$ and $B_\mathrm{1g}$ peak positions uniquely determines the sign of the tetragonal field $\Delta$, and our data firmly exclude $\Delta<0$ values (Supplemental Material~\cite{SM}). The obtained $\Delta/\lambda \simeq 0.4$ ratio is rather small, implying that all components of the orbital moment remain unquenched. While the relative strength of $\Delta$ and $\lambda$ is comparable to that in Sr$_2$IrO$_4$ with $\Delta/\lambda \simeq -0.5$~\cite{Kim14}, the signs of $\Delta$ in these compounds are opposite, despite similar lattice structures. The difference likely originates from screening of the lattice Madelung potential (which gives a large negative $\Delta$ in Sr$_2$IrO$_4$~\cite{Kat14}) in metallic Sr$_2$RhO$_4$, so that the ligand field $\Delta>0$ of the apically elongated (by $5\%$~\cite{Ito95}) octahedra dominates. It should be interesting to include such screening effects in quantum chemistry calculations~\cite{Kat14}.

The positive (negative) $\Delta$ values increase (reduce) the planar $xy$ orbital character of the effective $J=1/2$ wave function (Fig.~\ref{fig:3}). Consequently, the in-plane effective hopping amplitude in Sr$_2$RhO$_4$ (Sr$_2$IrO$_4$) is enhanced (suppressed) from its $\Delta=0$ value of $t_{eff}=\tfrac{2}{3}t$. This effect cooperates with the reduced spin-orbit coupling to support a metallic ground state in Sr$_2$RhO$_4$, and suggests that $c$-axis compression (changing the sign of $\Delta$) may trigger a metal-to-insulator transition.

Our results on the spin-orbit excitons with energy $\sim 230$ meV are complementary to ARPES data, which mostly probe coherent fermionic quasiparticles at and near the Fermi level \cite{Kim06,Bau06,San17,Bat20}. Results from both probes indicate a major influence of the spin-orbit coupling on the electronic structure and dynamics. Remarkably, however, the ARPES data suggest strong orbital polarization in favor of out-of-plane $xz/yz$ hole states, whereas the Raman data are more isotropic with preferential $xy$ character. The coexistence of spin-orbit excitons and fermionic quasiparticles with different orbital admixtures calls for a more elaborate description of our data, for instance via a dynamical mean-field theory approach to Raman scattering~\cite{Ble23}.

In summary, using Raman spectroscopy we have observed spin-orbit excitons in Sr$_2$RhO$_4$, as a direct signature of the spin-orbit entangled nature of correlated electrons in this Fermi-liquid metal. From the polarization dependence of the exciton peaks, we have quantified the spin-orbit and tetragonal crystal field parameters, which determine the orbital shapes of the corresponding wave functions. We found that the tetragonal field in Sr$_2$RhO$_4$ increases the planar $xy$ orbital character of the effective $J=1/2$ wave function, thereby supporting the metallic ground state. This finding calls for x-ray absorption and light-polarized ARPES studies of the orbital character of the electronic bands in Sr$_2$RhO$_4$. Our results also suggest that flattening of the RhO$_6$ octahedra may drive Sr$_2$RhO$_4$ into a Mott-insulating state, similar to the metal-insulator transition in Ca$_2$RuO$_4$~\cite{Bra98}. More generally, our results demonstrate that coherent fermionic quasiparticles and incoherent, atomiclike excitations can coexist in multiorbital metals, and that detailed analysis of the latter features can yield interesting insights into the microscopic interactions underlying the electronic structure. The experiments we have presented on a disorder-free system with a relatively simple electronic structure thus open up a potentially rich source of information on electronic correlations in other multiband metals.

We thank P. Puphal and C. Busch for technical help on crystal growth and characterization, and A. Schulz for technical assistance in Raman measurements. L.W. is supported by the Alexander von Humboldt foundation. H.L. acknowledges support by the W$\ddot{\rm u}$rzburg-Dresden Cluster of Excellence on Complexity and Topology in Quantum Matter \emph{ct.qmat} (EXC 2147, Project ID No.  390858490). We acknowledge financial support from the Deutsche Forschungsgemeinschaft (DFG, German Research Foundation), Project No. 492547816 TRR 360.

L.W. and H.L. contributed equally to this work.

\nocite{apsrev42Control}
\bibliographystyle{apsrev4-1}
%\bibliography{Reference_LW}

\clearpage
%\vspace{10cm}
\onecolumngrid

\begin{center}
\textbf{\large Supplemental Material for\\
Spin-orbit exciton in a correlated metal: Raman scattering study of Sr$_2$RhO$_4$}
\end{center}
\setcounter{equation}{0}
\setcounter{figure}{0}
\setcounter{table}{0}
\setcounter{page}{7}
\makeatletter
\renewcommand{\thesection}{S\Roman{section}}
\renewcommand{\thetable}{S\arabic{table}}
\renewcommand{\theequation}{S\arabic{equation}}
\renewcommand{\thefigure}{S\arabic{figure}}

\section{I. Additional experimental data}

The single-crystal samples were prepared with the optical floating-zone technique, then annealed in oxygen and characterized with magnetization and resistivity measurements to ensure complete oxygen stoichiometry. The magnetization was measured in a MPMS SQUID VSM with applied field of 1 T perpendicular to the $c$-axis of the crystal. The measurement was performed upon warming up after zero-field cooling. The resistivity was measured with the standard 4-point probe technique in PPMS with contacts attached to the side faces of the crystal for in-plane $\rho_{\rm ab}$ measurements, and on the top and bottom faces for $\rho_{\rm c}$ measurements. The results are presented in Fig. S1. The crystal orientation was determined by x-ray Laue diffraction, and confirmed by the phonon spectra shown in Fig. S2.

\begin{figure*}[h]
	\centering{\includegraphics[clip,width=16cm]{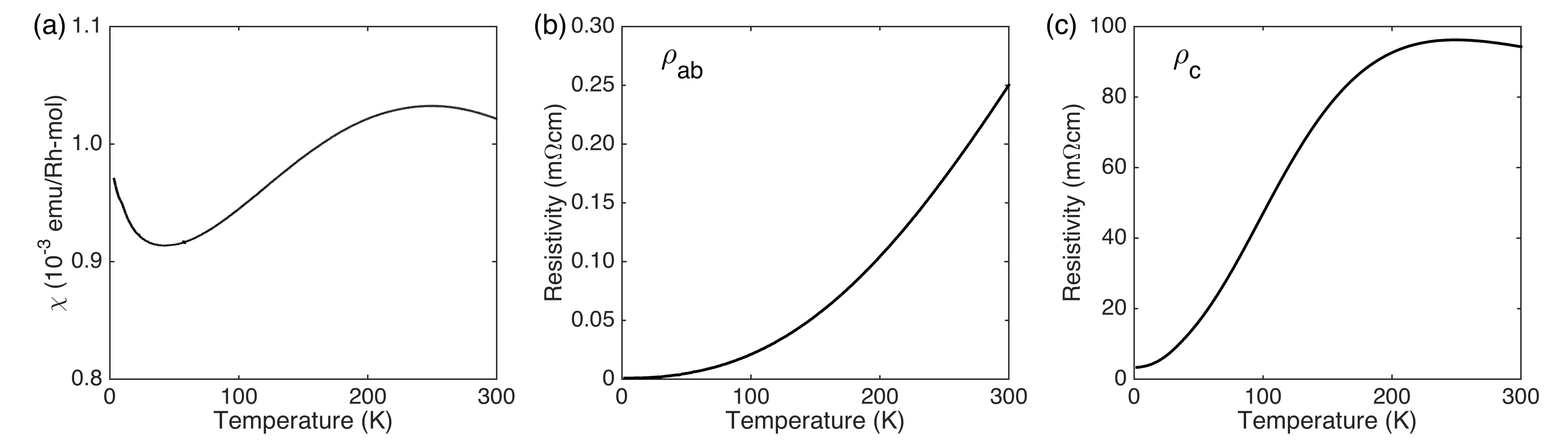}}
	\caption{Temperature dependence of (a) magnetization measured in a field of 1 T, and (b,c) in-plane $\rho_{\rm ab}$ and out-of-plane $\rho_{\rm c}$ resistivity. These results are consistent with those for full oxygen content samples~\cite{NagaiJPSJ2010}.}
       \label{fig1}
\end{figure*}

\begin{figure}[h]
	\centering{\includegraphics[clip,width=8.5cm]{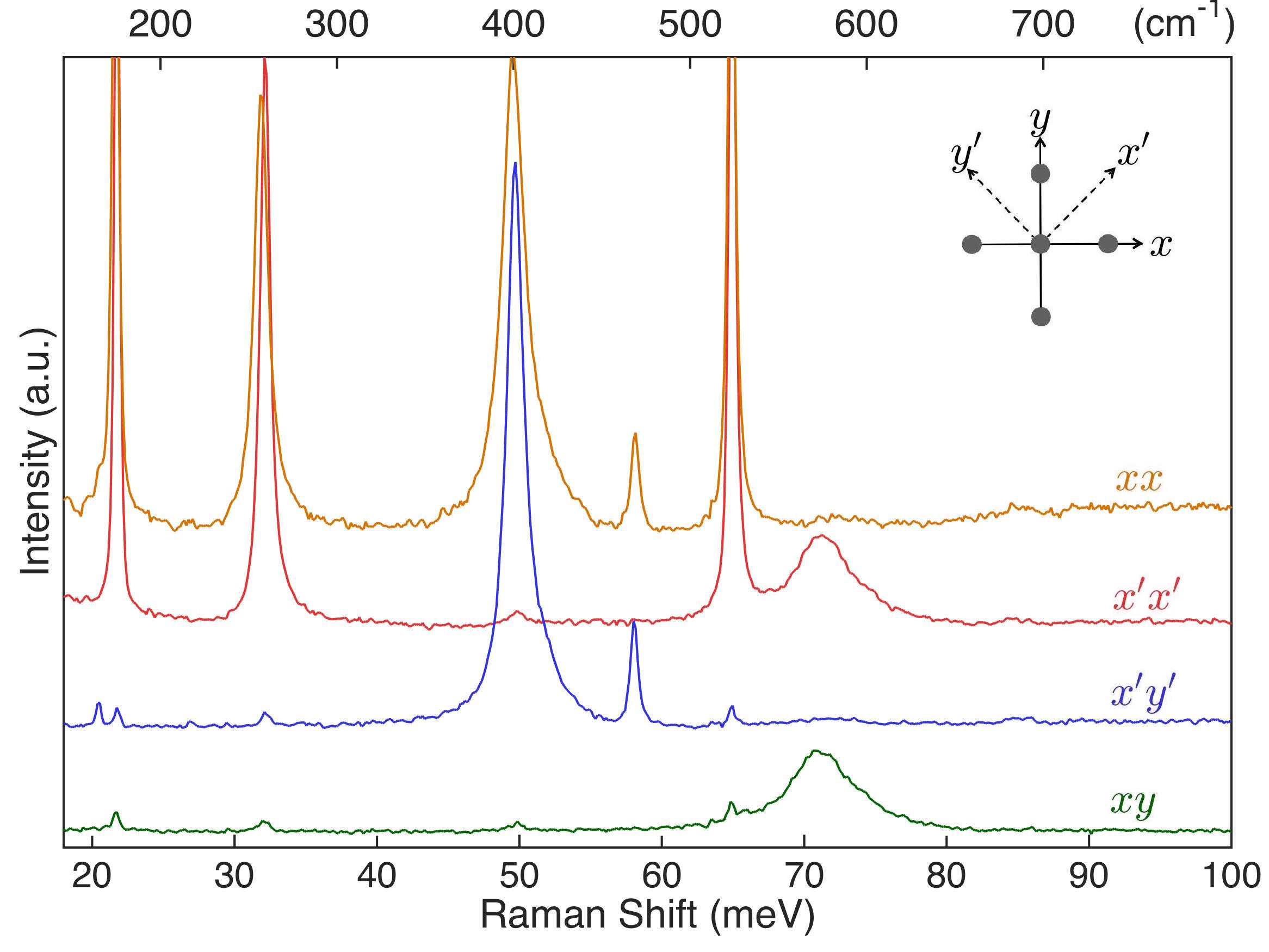}}
	\caption{Phonon spectra of Sr$_2$RhO$_4$ taken at 7 K. The data are vertically shifted for clarity. }
       \label{fig2}
     \end{figure}

During the Raman scattering measurements, the sample was mounted in a liquid-helium flow cryostat under ultrahigh vacuum, and studied in a confocal backscattering geometry using a Jobin Yvon LabRAM HR800 spectrometer. The laser beam with a power of $\sim$ 1 mW was focused to a $\sim$ 10-$\mu m$-diameter spot on the freshly cleaved surface, and the scattered photons were resolved by a 600 grooves/mm grating. Two samples were measured, and they yielded consistent results. Figure S3 shows the temperature dependence of Raman spectra for different linear photon polarizations. The spin-orbit exciton peaks persist in the $xx$, $x^\prime x^\prime$ and $x^\prime y^\prime$ channels from base to room temperature.

     \begin{figure*}[h]
	\centering{\includegraphics[clip,width=16cm]{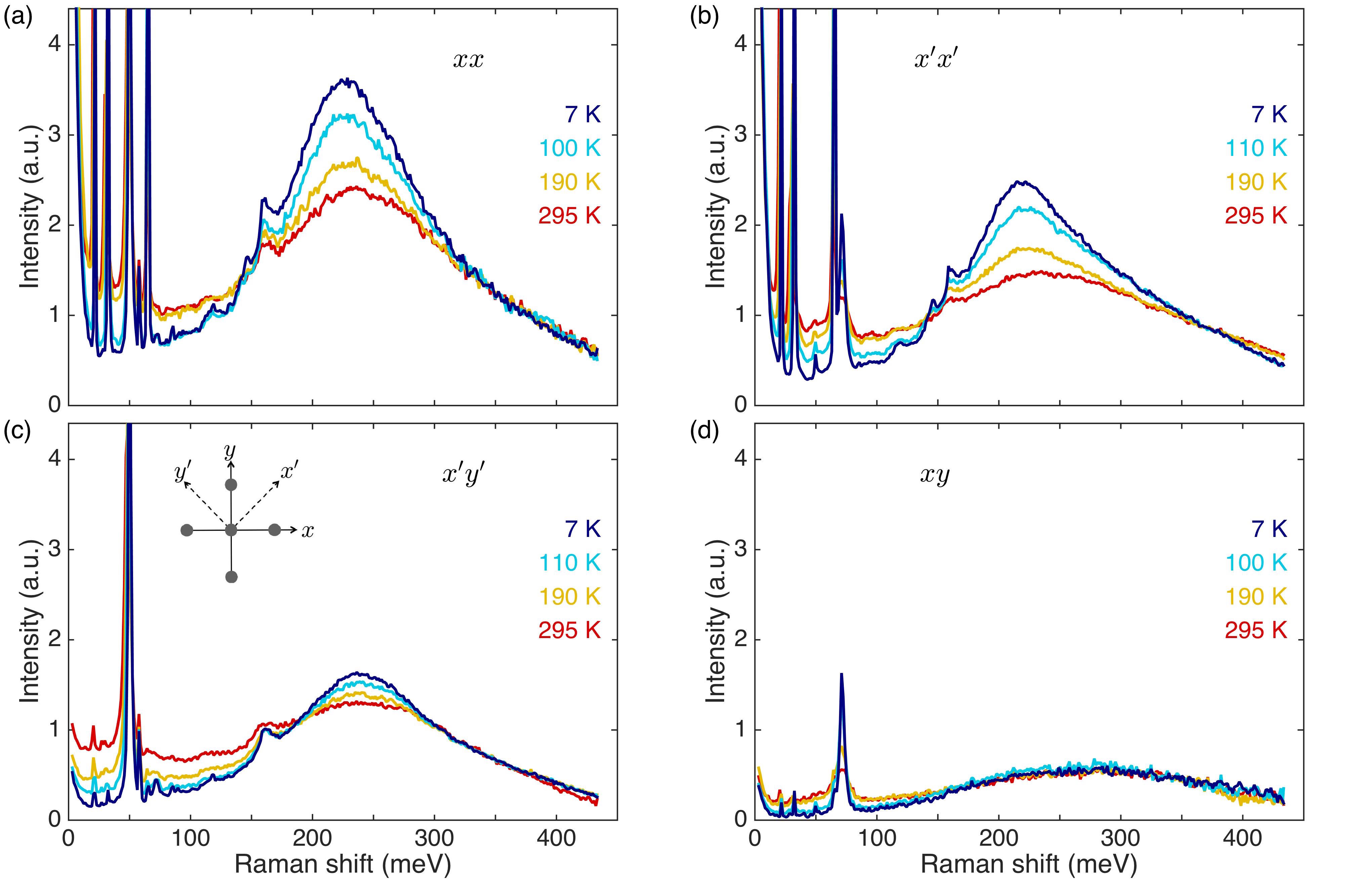}}
	\caption{Raman spectra taken at various temperatures in (a) $xx$, (b) $x^\prime x^\prime$, (c) $x^\prime y^\prime$, and (d) $xy$ polarizations.}
       \label{fig3}
     \end{figure*}

     The pure $A_\mathrm{1g}$ signal was extracted from the raw data (after removing the phonon peaks) in two different ways, as illustrated in Fig. S4. Both methods yield nearly identical spectra. Their average was taken as the resultant $A_\mathrm{1g}$ spectra and presented in Fig. 2 of the main text.

\begin{figure*}[h]
	\centering{\includegraphics[clip,width=8.0cm]{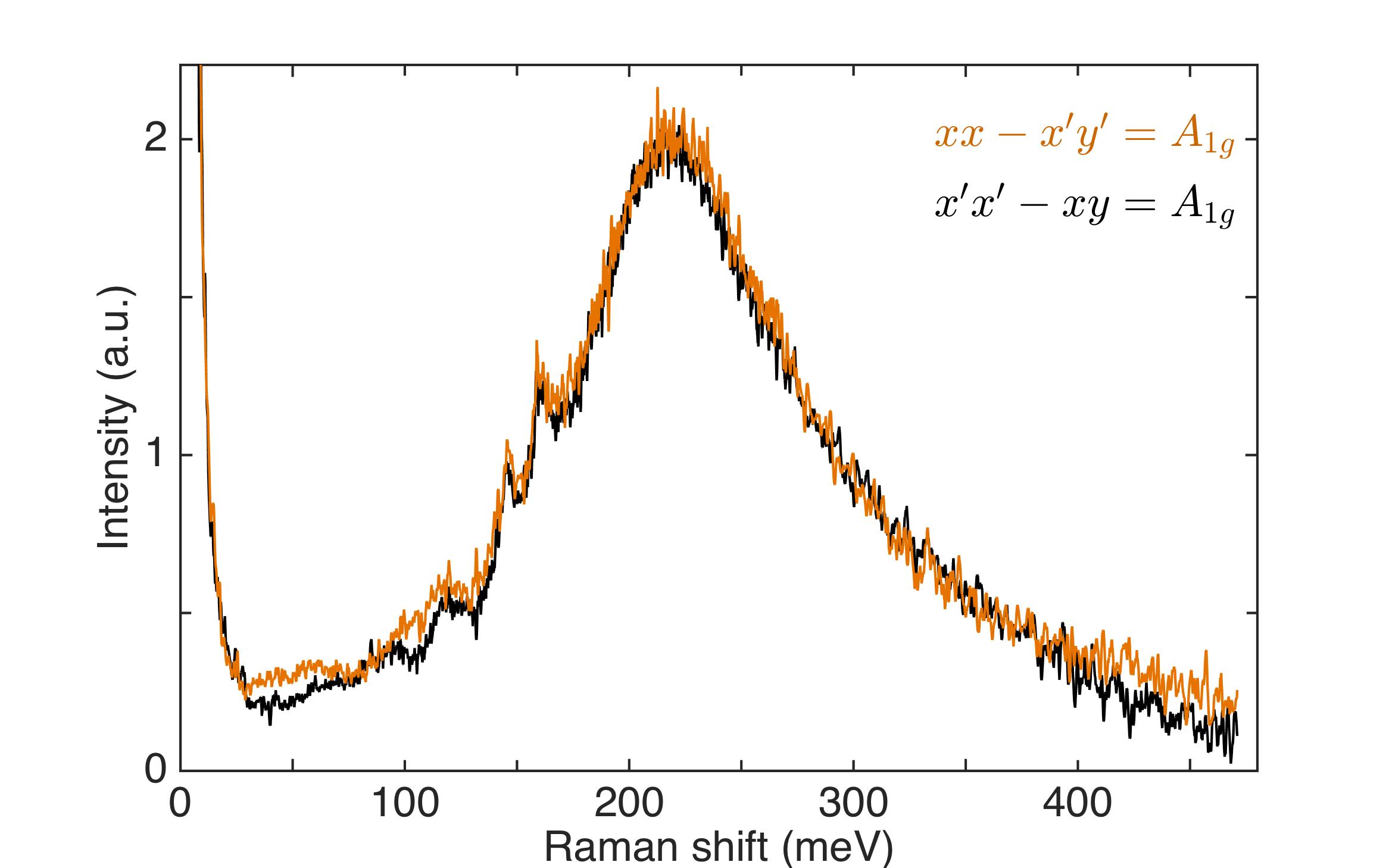}}
	\caption{The difference spectra between $xx$ and $x^\prime y^\prime$ polarizations (black), and between $x^\prime x^\prime$ and $xy$ polarizations (orange). Ideally, both should be identical to pure $A_\mathrm{1g}$ signal.}
       \label{fig4}
     \end{figure*}

\begin{figure*}[h]
	\centering{\includegraphics[clip,width=12.0cm]{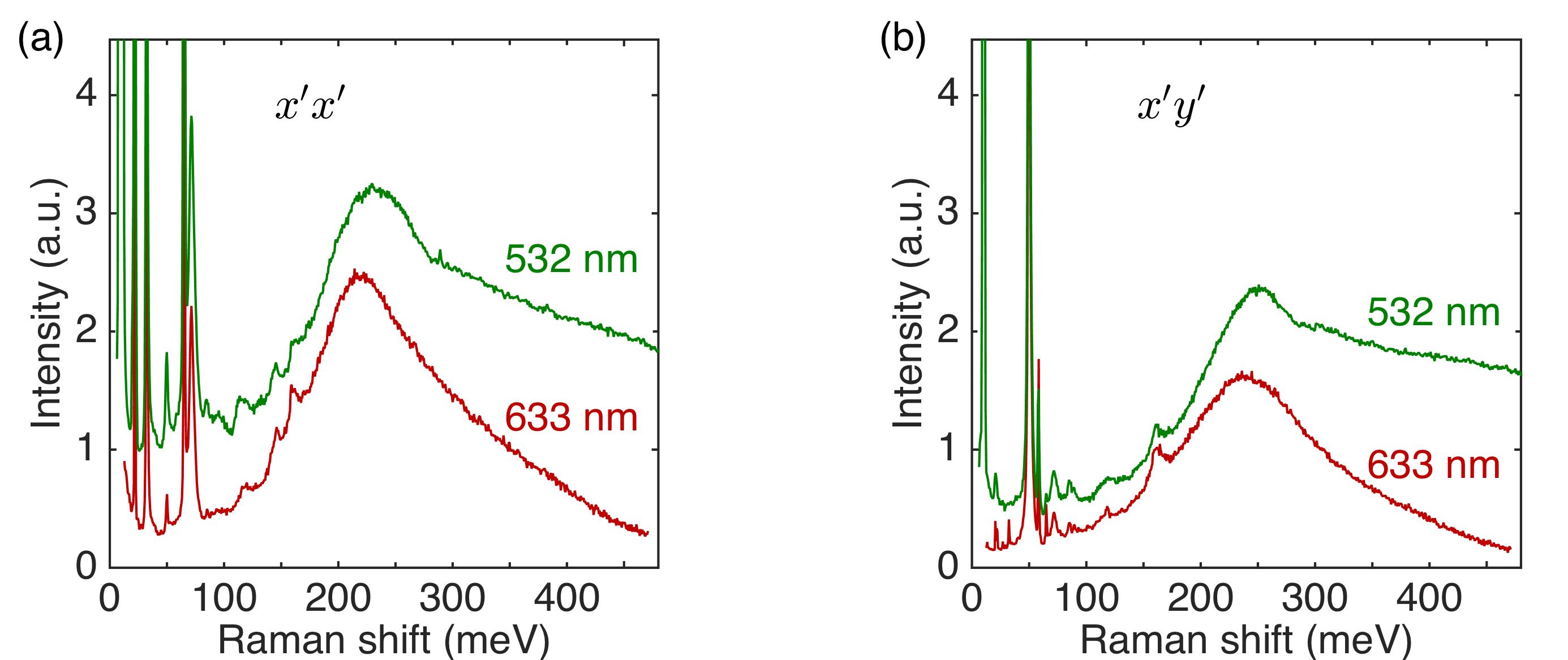}}
	\caption{Raman spectra taken at 7~K with different excitation lines.}
       \label{fig5}
     \end{figure*}

Most data were taken with the 632.8 nm (1.96 eV) excitation line from a He-Ne laser. To examine the influence of the incident photon energy, we also performed measurements with the 532 nm (2.33 eV) excitation line. As shown in Fig.~S5, the higher-energy 532 nm excitation line induces additional intensity in both polarization channels, especially at higher energies above $\sim$300 meV, which is likely due to fluorescence. The spin-orbit exciton peaks, however, are robust and well defined independent on the incident photon energy. Relative positions of the peaks at different channels are also similar, i.e. the $x^\prime x^\prime$ channel peak is lower in energy than the $x^\prime y^\prime$ channel by about 20 meV.

%---------------------------------------------------------------------------------------------
\section{II. Spin-orbital states of $t^5_{2g}$ electronic configuration in a tetragonal field}
%---------------------------------------------------------------------------------------------
The energy levels and wavefunctions shown in Fig.~3(c) of the main text are derived from the single-ion Hamiltonian, comprising the spin-orbit coupling $\lambda$ and tetragonal crystal field $\Delta$ terms:
\begin{align}
\mathcal {H} = \lambda (\bm{l \cdot s})+\frac{\Delta}{3}(n_{yz}+n_{zx}-2n_{xy}).
\label{eq:H}
\end{align}
We use here a hole representation, thus the positive $\Delta$ values stabilizing a planar $xy$ orbital would correspond -- within the point charge model -- to the case of elongated octahedra. It should be noticed, however, that the point charge model is often insufficient, so the actual sign of $\Delta$ in a real material has to be determined experimentally. In terms of effective orbital angular momentum states $|l_z\!=\!0\rangle\!=\!|xy\rangle$ and $|l_z\!=\!\pm1\rangle\!=\!-\frac{1}{\sqrt{2}}(i|zx\rangle\!\pm\!|yz\rangle)$, the tetragonal field term reads as $\Delta(l_z^2-\frac{2}{3})$. Diagonalization of this Hamiltonian results in three Kramers doublets: the ground state $A$ with eigenvalue
$\varepsilon_A=-\frac{1}{2}\varepsilon_B -\frac{1}{2}R$,
and the higher energy states $B$ and $C$ with
$\varepsilon_B=\frac{1}{2}\lambda+\frac{1}{3}\Delta$ and
$\varepsilon_C=-\frac{1}{2}\varepsilon_B +\frac{1}{2}R$, respectively. Here, $R= \sqrt{\frac{9}{4}\lambda^2+\Delta^2-\lambda\Delta}$. The excitation energies corresponding to the $A\rightarrow B$ and $A\rightarrow C$ transitions then follow as:
 \begin{align}
E_\mathrm{B}&=\varepsilon_B-\varepsilon_A=
               \frac{3}{4}\lambda+\frac{1}{2}\Delta +\frac{1}{2}R\;, \notag \\
E_\mathrm{C}&=\varepsilon_C-\varepsilon_A=R \;.
 \end{align}
At $\Delta=0$, one has $R=\frac{3}{2}\lambda$, and thus $E_C=E_C=\frac{3}{2}\lambda$.

To calculate the spectral weights of spin-orbit excitons in the Raman spectra, we need to know the wavefunctions $|A\rangle$ etc. In the basis of $|l_z,s_z\rangle$ states, the ground state Kramers doublet functions read as:
\begin{align}
|A_+\rangle&= s_{\theta}\;|0,+\tfrac{1}{2}\rangle-
c_{\theta}  \;|\!+1,-\tfrac{1}{2}\rangle, \notag \\
|A_-\rangle&= s_{\theta} \;|0,-\tfrac{1}{2}\rangle-
c_{\theta} \;|\!-1,+\tfrac{1}{2}\rangle.
\label{eq:wf1}
\end{align}
Here $c_\theta \equiv \cos\theta$, $s_\theta \equiv \sin\theta$, and angle $0\leq\theta\leq\frac{\pi}{2}$ is given by $\tan 2\theta = 2\sqrt{2}\lambda/(\lambda-2\Delta)$. In the cubic limit of $\Delta=0$ with $s_\theta=\sqrt{\frac{1}{3}}$ and $c_\theta=\sqrt{\frac{2}{3}}$, Eq.~\ref{eq:wf1} represents ideal total angular momentum $J=\frac{1}{2}$ states, equally contributed by all three $t_{2g}$ orbitals. At large positive (negative) $\Delta$ values, the angle $\theta$ approaches $90^\circ$ ($0^\circ$), and the $xy$ orbital contribution $\propto s_\theta$ to the ground state is enhanced (suppressed).

The $B$ and $C$ doublets correspond to the $J_z=\pm\frac{3}{2}$ and $J_z=\pm\frac{1}{2}$ states of $J=\frac{3}{2}$ quartet, respectively:
\begin{align}
 |B_+\rangle& = |\!-1,-\tfrac{1}{2}\rangle,\ \ \ \ \ \ \ &|C_+\rangle= c_{\theta} \;|0,+\tfrac{1}{2}\rangle+s_{\theta} \;|\!+1,-\tfrac{1}{2}\rangle, \notag\\
 |B_-\rangle& = |\!+1,+\tfrac{1}{2}\rangle. \ \ \ \ \ \ \ &|C_-\rangle= c_{\theta} \;|0,-\tfrac{1}{2}\rangle+s_{\theta} \;|\!-1,+\tfrac{1}{2}\rangle.
\label{eq:wf2}
\end{align}

To describe the transitions between different spin-orbital states, we now introduce the following composite operators. First of all, we define the raising $\sigma^+$ and lowering $\sigma^-$ operators, such that they change the total angular momentum by $\Delta J_z  =1$ and $\Delta J_z  =-1$, correspondingly:
\begin{alignat}{3}
\sigma^+&=A_+^{\dag} A_- \;, \ \ \ \ \ \ \ \ \ & \sigma^+_{BA}&=B_-^{\dag} A_+ \;, \ \ \ \ \  \ \ \ \ & \sigma^+_{CA}&=C_+^{\dag} A_- \;,
\notag \\
\sigma^-&=A_-^{\dag} A_+ \;, \ \ \ \ \ \ \  \ \ & \sigma^-_{BA}&=B_+^{\dag} A_- \;, \ \ \  \ \ \ \ \ \ & \sigma^-_{CA}&=C_-^{\dag} A_+ \;.
\end{alignat}
Accordingly, the transverse components $\sigma^x=(\sigma^+ + \sigma^-)$, $\sigma^y= (\sigma^+ - \sigma^-)/i$ of the $A$-doublet pseudospin, and similar operators corresponding to the exciton transitions $\sigma_{BA}^x=(\sigma_{BA}^+ + \sigma_{BA}^-)$, $\sigma_{BA}^y=(\sigma_{BA}^+ - \sigma_{BA}^-)/i$, and $\sigma_{CA}^x=(\sigma_{CA}^+ + \sigma_{CA}^-)$, $\sigma_{CA}^y=(\sigma_{CA}^+ - \sigma_{CA}^-)/i$ change the angular momentum by $\Delta J_z=\pm 1$.  

From the wavefunctions given by  Eqs.~\eqref{eq:wf1} and \eqref{eq:wf2}, it also follows that the $A$-doublet density $n=(A_+^{\dag} A_+ + A_-^{\dag} A_-)$ and its pseudospin $z$-component $\sigma^z=(A_+^{\dag} A_+ - A_-^{\dag} A_-)$, as well as similar operators in the $C$ exciton channel $n_{CA}=(C_+^{\dag} A_+ + C_-^{\dag} A_-)$ and $\sigma_{CA}^z=(C_+^{\dag} A_+ - C_-^{\dag} A_-)$ preserve the angular momentum projection, i.e. $\Delta J_z = 0$. However, the corresponding $B$ exciton operators $n_{BA}=(B_+^{\dag} A_+ + B_-^{\dag} A_-)$ and $\sigma_{BA}^z=(B_+^{\dag} A_+ - B_-^{\dag} A_-)$ change $J_z$ by $\Delta J_z = \pm 2$. The difference in the selection rules for $A\!\rightarrow\!B$ and $A\!\rightarrow\!C$ exciton transitions is due to that the $B$ and $C$ states have different $J_z$ values and thus transform differently under $C_4$ rotation. The operators introduced here and their distinct selection rules will be useful for the symmetry analysis of the Raman scattering operators below.     

%--------------------------------------------------------------------------------------------------------
\section{III. Raman operators and spin-orbit exciton intensities}
%--------------------------------------------------------------------------------------------------------

The processes describing Raman light scattering on spin-orbit excitons are illustrated in Fig.~4 of the main text. The corresponding operators and their matrix elements can in principle be obtained by a direct consideration of the  intersite dipolar transitions between various spin-orbital states involved in these processes. Alternatively, we can follow the Fleury-Loudon approach~\cite{Fle1968,Dev2007} where one expresses the Raman scattering operator in terms of the electron exchange Hamiltonian $\mathcal H_{ij}$: 
\begin{align}
\mathcal R(\vc \epsilon, \vc \epsilon') 
\propto \sum_{\langle ij\rangle} (\vc \epsilon\cdot \vc r_{ij}) (\vc \epsilon' \cdot \vc r_{ij}) \mathcal H_{ij}.
\label{eq:Rop}
\end{align}
Here, $\vc \epsilon$ and $\vc \epsilon'$ are the incident and scattered photon polarization vectors (both are in the $xy$ plane in our experiment), respectively, and summation is over the exchange pairs $\langle ij\rangle$. In the present multiorbital case, the exchange Hamiltonian $\mathcal H_{ij}$ contains all the transitions between various  spin and orbital states, including the processes that create the spin-orbit excitons of our interest. The explicit form of $\mathcal H_{ij}$ for $t_{2g}$ orbitals and spins $s=1/2$ in the perovskite lattice is well known~\cite{Kha2003,Kha2005}; for the nearest-neighbor (NN) bonds along the $x$-axis it reads as   
\begin{align}
\mathcal{H}_{ij}^{(x)} =\mathcal{J} \left[ \left(4 \vc s_i \cdot \vc s_j+ 1 \right)\mathcal{O}^{(x)}_{ij}
+n_{i,yz}+n_{j,yz}  - \tau^2 (n_{i,xy}+n_{j,xy}) \right].
\label{eq:Hxn}
\end{align}
Here, $\mathcal{O}^{(x)}_{ij} = n_{i,zx}n_{j,zx} \!+\! n_{i,xy}n_{j,xy} \!+\!(d_{zx}^{\dag}d_{xy})_i (d_{xy}^{\dag}d_{zx})_j \!+\! (d_{xy}^{\dag}d_{zx})_i (d_{zx}^{\dag}d_{xy})_j$, $n_{xy} \!=\! d_{yz}^{\dag}d_{yz}$, and $\tau \!=\! \sin 2 \alpha (t_{pd\sigma} / t_{pd\pi})$, where $\alpha$ measures the deviation of the Rh-O-Rh bonding angle from 180$^\circ$ (see Fig.~4 of the main text). $\mathcal{J}=t^2/U$ stands for $t_{2g}$ orbital exchange constant, where $t$ denotes the NN hopping integral and $U$ is Hubbard repulsion. In spin-orbit coupled systems, it is more convenient to represent $\mathcal H_{ij}$ in terms of the orbital momentum  operators $l_x = i (d_{xy}^{\dag}d_{zx} - d_{zx}^{\dag}d_{xy})$, $l_y = i (d_{yz}^{\dag}d_{xy} - d_{xy}^{\dag}d_{yz})$, and $l_z = i (d_{zx}^{\dag}d_{yz} - d_{yz}^{\dag}d_{zx})$: 
\begin{align}
\mathcal{H}_{ij}^{(x)} =\mathcal{J} \left[ \left(4 \vc s_i \cdot \vc s_j+ 1 \right)\mathcal{O}^{(x)}_{ij}
-l^2_{xi}-l^2_{xj} + \tau^2 (l^2_{zi}+l^2_{zj}) \right] \;,
\label{eq:Hxl}
\end{align}
where $\mathcal{O}^{(x)}_{ij}= [(1-l^2_y)_i(1-l^2_y)_j+(l_y l_z)_i(l_z l_y)_j]+[y\leftrightarrow z]$. The Hamiltonian $\mathcal{H}_{ij}^{(y)}$ for $y$-type bonds is obtained by substitution $l_y\rightarrow l_x$. Summation of $\mathcal{H}_{ij}^{(\gamma)}$ over the $\gamma=x$ or $y$ bonds gives $\mathcal{R}_\gamma$ operator in Eq.~(2) of the main text. 

In general, the operators in Eq.~\eqref{eq:Hxl} create all possible transitions between the spin-orbit entangled $A,B$, and $C$ states defined by Eqs.~\eqref{eq:wf1}-\eqref{eq:wf2} above. This results in a number of terms corresponding to Raman scattering from ({\bf a}) low energy pseudospin fluctuations within the ground state $A$ doublets (resulting in two-magnon scattering in insulators), ({\bf b}) spin-orbit exciton transitions $A \!\rightarrow \! B,C$ on one site of the exchange $\langle ij\rangle$ pair (corresponding to the processes in Fig.~4 of the main text), and ({\bf c}) spin-orbit excitons on both sites $i$ and $j$. We recall that the spin-orbit exciton is created by intersite hopping between $J=1/2$ and $J=3/2$ states. As this hopping is small in perovskites with $180^\circ$ Me-O-Me bonding geometry, we keep only single-exciton terms, which are essential to describe our data. This gives, for the NN exchange bonds $\langle ij\rangle_x$ and $\langle ij\rangle_y$ along $x$ and $y$ directions of a square lattice, the following result:
\begin{align}
\mathcal{R}_x &\!=\!\tfrac{1}{4} c_{\theta} \!\sum_{i,j_x} \left[c^2_{\theta}(+n_{BA}\!-\!s_{\theta}n_{CA})_i n_j
  + (1\!+\!s^2_{\theta}) ( s_{\theta} \vc \sigma_{CA} \vc \sigma + \sigma^y_{BA}  \sigma^y \!-\! \sigma^z_{BA}  \sigma^z)_{ij} \!-\!c^2_{\theta}(\sigma^x_{BA})_i \sigma^x_j \!-\! 4\tau^2s_\theta(n_{CA})_i\right] +H.c.,
\notag \\
\mathcal{R}_y &\!=\! \tfrac{1}{4} c_{\theta} \!\sum_{i,j_y}\!\left[c^2_{\theta}(-n_{BA}\!-\!s_{\theta}n_{CA})_i n_j
 + (1\!+\!s^2_{\theta})( s_{\theta} \vc \sigma_{CA} \vc \sigma \!+\! \sigma^x_{BA}  \sigma^x \!+\! \sigma^z_{BA}  \sigma^z)_{ij} - c^2_{\theta}  (\sigma^y_{BA})_i \sigma^y_j \!-\! 4\tau^2s_\theta (n_{CA})_i\right]+H.c.,
\label{eq:soe}
\end{align}
 Here, the exciton transitions $A\!\rightarrow\!B$ and $A\!\rightarrow\!C$ are created on site $i$, while a hole on nearest-neighbor sites $j=i \pm x$ or $j=i \pm y$ remains within the ground state $A$ doublet, as illustrated in  Fig.~4 of the main text. From the selection rules for exciton operators $n_{BA}$, $n_{CA}$, etc., which have been specified in the previous section, it follows that $\mathcal{R}_x$ and $\mathcal{R}_y$ operators change the angular momentum projection $J_z$ by $0$ or $\pm 2$, and under $C_4$ rotation around the $z$ axis they transform into each other, as expected. 

The Raman operators $\mathcal{R}_{A_{\rm 1g}}$ and $\mathcal{R}_{B_{\rm 1g}}$ are given by $\mathcal{R}_x \pm \mathcal{R}_y$. It is convenient to introduce the combinations $n^i_s=\tfrac{1}{4}(n_{i+x}+n_{i-x}+n_{i+y}+n_{i-y})$ and $n^i_d=\tfrac{1}{4}(n_{i+x}+n_{i-x}-n_{i+y}-n_{i-y})$ of $A_{\rm 1g}$ and $B_{\rm 1g}$ type symmetries, correspondingly. Then, the non-spin related part of the Raman operator can be written as
\begin{align}
\mathcal{R}^n_{A_{\rm 1g}} =c_{\theta}\sum_i \left[ c^2_{\theta}\; n_dn_{BA} - s_{\theta} (c^2_{\theta}\; n_s + 4\tau^2) n_{CA} \right]_i+H.c.\;, \ \ \ \ \ \
  \mathcal{R}^n_{B_{\rm 1g}} = c^3_{\theta}\sum_i\left[n_sn_{BA} - s_{\theta}\; n_dn_{CA}\right]_i+H.c.\;.
  \label{eq:Rn}
\end{align}
We recall that $B$ and $C$ states have different $J_z$ values, see Eq.~\eqref{eq:wf2}, thus the corresponding operators $n_{BA}$ and $n_{CA}$ transform differently under $\pi/2$ rotations about the $z$ axis, i.e. with and without the sign change, respectively. This guarantees correct symmetry relations in the above equations (e.g., operator $n_dn_{BA}$ transforms as an object of $A_{\rm 1g}$ symmetry).

Similarly, we introduce the $A$ doublet pseudospin combinations $\vc{\sigma}^i_{s/d}=\tfrac{1}{2}(\vc{\sigma}_{i+x}+\vc{\sigma}_{i-x}\pm \vc{\sigma}_{i+y}\pm \vc{\sigma}_{i-y})$ near the central site $i$, and find the spin-related part of the Raman operator in the following form:
\begin{align}
\mathcal{R}^\sigma_{A_{\rm 1g}} = \tfrac{1}{2}& c_\theta \sum_i\left[
s_\theta (1+s^2_\theta)\; \vc \sigma_s \vc \sigma_{CA} -(1+s^2_\theta)\;
\sigma^z_d \sigma^z_{BA}  - (\sigma^x_d \sigma^x_{BA} - \sigma^y_d \sigma^y_{BA}) +
s^2_\theta (\sigma^x_s\sigma^x_{BA} + \sigma^y_s\sigma^y_{BA})\right]_i+H.c. ,
                                                   \notag \\
\mathcal{R}^\sigma_{B_{\rm 1g}} = \tfrac{1}{2}& c_\theta \sum_i\left[
s_\theta (1+s^2_\theta)\; \vc \sigma_d \vc \sigma_{CA} -(1+s^2_\theta)\;
\sigma^z_s \sigma^z_{BA} - (\sigma^x_s \sigma^x_{BA} - \sigma^y_s \sigma^y_{BA}) +
s^2_\theta  (\sigma^x_d\sigma^x_{BA} + \sigma^y_d\sigma^y_{BA})\right]_i+H.c. .
\label{eq:Rs}
\end{align}
Here, exciton creation on site $i$ is accompanied by the $A$-doublet pseudospin dynamics on the nearest-neighbor sites. Again, using the above selection rules for $\sigma_{BA}^z$, etc., one can verify that under $C_4$ rotation around the $z$ axis, $\mathcal{R}^\sigma$ operators in Eq.~\eqref{eq:Rs} transform according to the $A_{\rm 1g}$ and $B_{\rm 1g}$ symmetry rules, respectively.

Calculation of the Raman lineshapes, which are determined by the correlation functions of the $\mathcal{R}^n$ and $\mathcal{R}^\sigma$ operators, is obviously a nontrivial task; it is, however, rather straightforward to evaluate the spectral weights of the exciton transitions. In the density channel, we approximate the hole density $n$ in the ground state $A$ doublet as $\langle n \rangle \simeq 1$. This gives $n_s\simeq 1$ and $n_d\simeq 0$, and from Eq.~\eqref{eq:Rn} we find the intensities of the $A \rightarrow B$ and $A \rightarrow C$ transitions:
\begin{alignat} {3}
 & \ \ \ \ \ \ \ \ &  \ \ \ \ \ \ \ \ & B \ \ {\rm exciton}  &\ \ \ \ \ \ \ \ \ \ \ \ \ \ \ \  & C \ \ {\rm exciton}
\notag \\
&I^n_{A_{\rm 1g}}: &      &0& &s^2_\theta c^2_\theta(c^2_\theta+4\tau^2)^2
\notag \\
&I^n_{B_{\rm 1g}}: &      &c^6_\theta & &0 \;.
\label{eq:a1}
\end{alignat}
We recall that $\tau=\tilde{t}/t = \sin 2\alpha \;(t_{pd\sigma}/t_{pd \pi})$,  where $\tilde{t}$ is the hopping amplitude between $t_{\rm 2g}$ and $e_g$ orbitals, which is allowed due to in-plane rotations of the RhO$_6$ octahedra. In Sr$_2$RhO$_4$, the angle $\alpha=10.5^{\circ}$~\cite{Sub1994} and $\sin 2\alpha=0.36$.

For the transitions involving spin dynamics, we neglect non-local spin correlations (a reasonable approximation for paramagnetic Sr$_2$RhO$_4$), by setting $\langle\sigma^\alpha_i \sigma^\beta_j \rangle = \delta_{i,j} \delta_{\alpha,\beta}$. This gives $\langle\sigma^\alpha_s \sigma^\beta_s \rangle_i = \langle\sigma^\alpha_d \sigma^\beta_d \rangle_i = \delta_{\alpha,\beta}$, while $\langle\sigma^\alpha_s \sigma^\beta_d \rangle_i =0$ by symmetry. As a result, we find
\begin{align}
& B \ \ {\rm exciton}  &\ \ \ \ \ \  & C \ \ {\rm exciton}
\notag \\
I^\sigma_{A_{\rm 1g}} = I^\sigma_{B_{\rm 1g}}: \ \ \ \ \ \ \ \ \ \ \ \ \ &\tfrac{1}{4}c^6_\theta + \tfrac{1}{2} c^2_\theta (1+s^2_\theta)^2 & &\tfrac{3}{4} c^2_\theta s^2_\theta (1+s^2_\theta)^2\;.
\label{eq:a2}
\end{align}
From Eqs.~\eqref{eq:a1} and \eqref{eq:a2}, the total intensities $I_{B/C}=(I^n+I^\sigma)_{B/C}$ in Eqs.~(3) and (4) of the main text, follow.

\begin{figure*}[h]
\centering{\includegraphics[clip,width=18 cm]{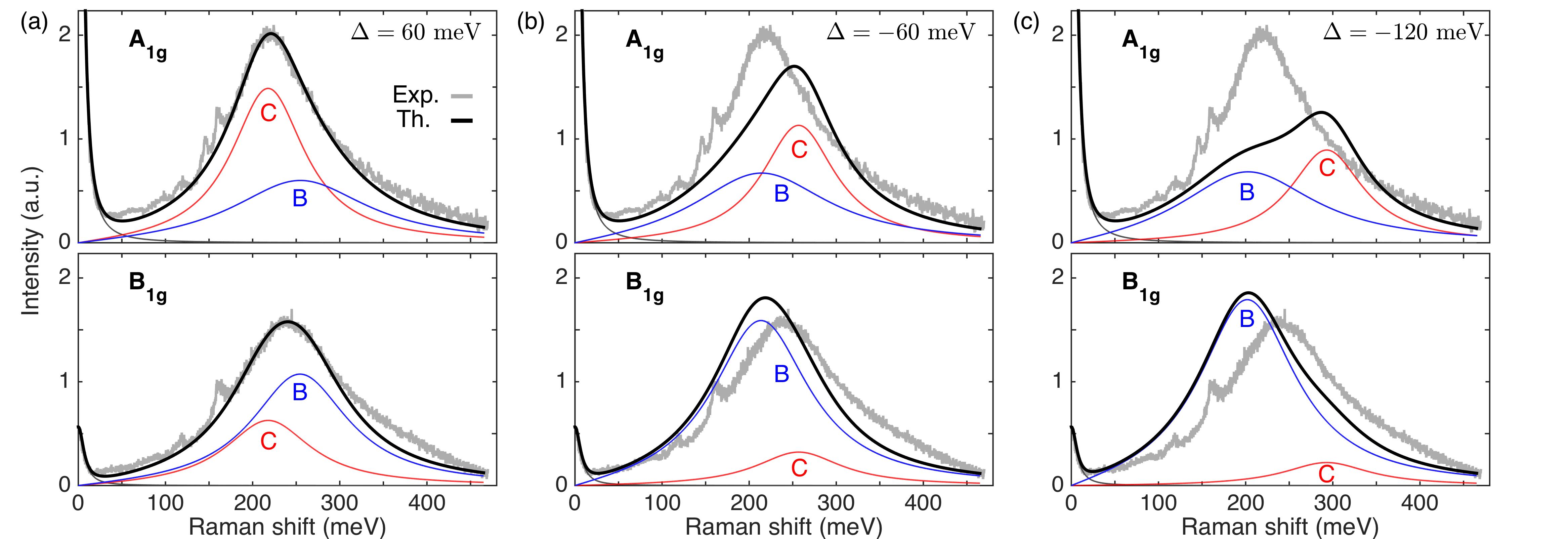}}
\caption{Raman spectra calculated using different $\Delta$ values in Eq.~\eqref{eq:H}: (a) 60 meV, (b) -60 meV, and (c) -120 meV. Other parameters used are: $\lambda$=154 meV and $t_{pd\sigma}/t_{pd \pi}=1.2$. The notations and experimental data are the same as in Fig.~2 of the main text.}
\label{fig6}
\end{figure*}

A few comments on the above results are in order. First, both $B$ and $C$ excitons contribute to $A_{\rm 1g}$ and $B_{\rm 1g}$ spectra. Their relative spectral weights in the two channels are different, however, with the $C$ and $B$ contributions being more pronounced in $A_{\rm 1g}$ and $B_{\rm 1g}$ spectra, respectively. This allows one to determine the crystal field parameter $\Delta$ from the polarization dependence of the Raman spectra. As illustrated in Fig.~\ref{fig6}, the $A_{\rm 1g}$ and $B_{\rm 1g}$ peaks switch their relative positions as $\Delta$ changes its sign, and the experimental data clearly rules out negative $\Delta$ values in Sr$_2$RhO$_4$.

%--------------------------------------------------------------------------

%--------------------------------------------------------------------------
\begin{figure*}[h]
\centering{\includegraphics[clip,width=17 cm]{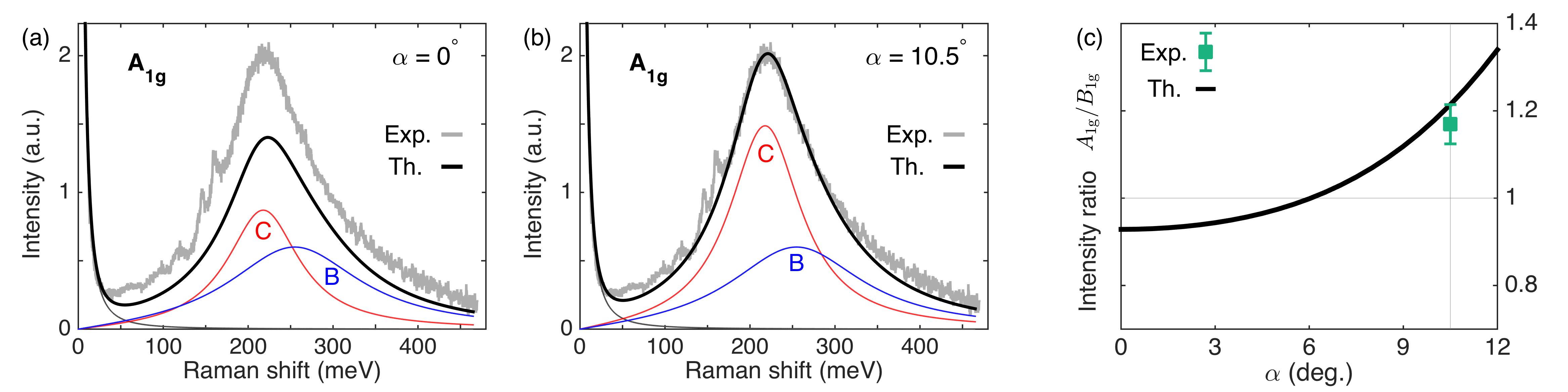}}
\caption{The $A_{\rm 1g}$ Raman spectra calculated using different octahedral rotation angles: (a) $\alpha=0$, and (b) $\alpha=10.5^{\circ}$ as in Sr$_2$RhO$_4$~\cite{Sub1994}. (c) Intensity ratio between $A_{\rm 1g}$ and $B_{\rm 1g}$ channels as a function of angle $\alpha$. The experimental ratio is reproduced at a realistic value of $\alpha=10.5^{\circ}$. The parameters used are: $\lambda$=154 meV, $\Delta$=60 meV, and $t_{pd\sigma}/t_{pd \pi}=1.2$.}
\label{fig7}
\end{figure*}
%--------------------------------------------------------------------------

Second, we note that calculations within the $t_{2g}$ orbital space alone, i.e. without including $\tilde t$ hopping to the $e_g$ states, give the $A_{\rm 1g}$ intensity always smaller than that in the $B_{\rm 1g}$ channel, independent of the $\Delta/\lambda$ ratio. In fact, as an attempt to remedy this discrepancy with experiment, we first inspected the Raman process within a single RhO$_6$ unit cell. Due to spin-orbit coupling, local $A \rightarrow B,C$ excitations are possible via O$(2p)$-Rh$(4d)$ dipolar transitions. Simple analysis of this scenario gives $I_{A_{\rm 1g}}\propto s_\theta^2c_\theta^2$ and $I_{B_{\rm 1g}}\propto c_\theta^2$, i.e. exciton spectral weight in the $A_{\rm 1g}$ channel would have been smaller by a factor of $s_\theta^2 \sim 1/3$, in sharp contrast to experimental result of $I_{A_{1g}}>I_{B_{1g}}$. This observation has prompted us to include the $\tilde t$-hopping processes, which are activated by the rotations of the RhO$_6$ octahedra. Note that these processes contribute only to the $A_{\rm 1g}$ channel, see Eq.~\eqref{eq:a1}. As a result, the $A_{\rm 1g}$ intensity is enhanced and the $A_{\rm 1g}$ vs $B_{\rm 1g}$ intensity issue is resolved, as demonstrated in Fig.~\ref{fig7}.

Third, we assumed above that the light induced dipolar transitions are dominated by the NN intersite hoppings, as it is in perovskite lattices with nearly $180^\circ$ Me-O-Me bonding geometry. Within this approximation, Eq.~\eqref{eq:Rop} suggests no Raman scattering intensity in the $B_{2g}$ channel. However, $d$  electron hoppings $t'$ beyond the NN bonds are always present (typically of the order of $t' \sim 0.2 t$), and thus a small but finite $B_{2g}$ intensity is expected. 

Finally, we note that the above theory -- with modifications including magnetic order effects -- is applicable to Sr$_2$IrO$_4$, which has a very similar lattice structure. The single-exciton peaks are also observed in this material~\cite{YangPRB2015}. Importantly, the relative order of the $A_{\rm 1g}$ and $B_{\rm 1g}$ peak positions ($A_{\rm 1g}$ is higher in energy than $B_{\rm 1g}$) is different from that in Sr$_2$RhO$_4$. We also note that, in contrast to perovskites, the intersite hopping between $J=1/2$ and $J=3/2$ states is not small in compounds with $90^\circ$ Me-O-Me bonding geometry, making the double-exciton transitions equally important for the Raman scattering process. A nice example of this is the Kitaev material RuCl$_3$, where both single and double-exciton peaks have been clearly observed~\cite{WarzanowskiPRR2020,Lee2021multiple}.

\end{document}